\documentclass[aps,pra,superscriptaddress,twocolumn,showpacs,longbibliography,titlepage=false,floatfix]{revtex4-2}
\usepackage{amsmath, amsthm, amssymb, graphicx, txfonts}
\usepackage{xcolor}
\usepackage{nicefrac}
\usepackage[breaklinks=true,colorlinks=true,linkcolor=blue,urlcolor=blue,citecolor=blue]{hyperref}
\begin{document}

\newcommand{\reals}{\mathbb{R}}
\newcommand{\complex}{\mathbb{C}}
\newcommand{\expect}[1]{\ensuremath{\left\langle#1\right\rangle}}
\newcommand{\ket}[1]{\ensuremath{\left|#1\right\rangle}}
\newcommand{\bra}[1]{\ensuremath{\left\langle#1\right|}}
\newcommand{\braket}[2]{\ensuremath{\left\langle#1|#2\right\rangle}}
\newcommand{\ketbra}[2]{\ket{#1}\!\!\bra{#2}}
\newcommand{\braopket}[3]{\ensuremath{\bra{#1}#2\ket{#3}}}
\newcommand{\proj}[1]{\ketbra{#1}{#1}}
\newcommand{\rrangle}{\rangle\!\rangle} \newcommand{\llangle}{\langle\!\langle}
\newcommand{\sket}[1]{\ensuremath{\left|#1\right\rrangle}}
\newcommand{\sbra}[1]{\ensuremath{\left\llangle#1\right|}}
\newcommand{\sbraket}[2]{\ensuremath{\left\llangle#1|#2\right\rrangle}}
\newcommand{\sketbra}[2]{\sket{#1}\!\!\sbra{#2}}
\newcommand{\sbraopket}[3]{\ensuremath{\sbra{#1}#2\sket{#3}}}
\newcommand{\sproj}[1]{\sketbra{#1}{#1}}
\def\Id{1\!\mathrm{l}}
\newcommand{\Tr}{\mathrm{Tr}}
\newcommand{\cM}{\mathcal{M}}
\newcommand{\cP}{\mathcal{P}}
\newcommand{\cG}{\mathcal{G}}
\newcommand{\cD}{\mathcal{D}}
\newcommand{\cE}{\mathcal{E}}
\newcommand{\cL}{\mathcal{L}}
\newcommand{\cU}{\mathcal{U}}
\newcommand{\cH}{\mathcal{H}}

\newcommand{\note}[1]{{\color{red}#1}}
\newcommand{\model}{{\mathcal{M}}}
\newcommand{\wildcard}{{\mathcal{W}}}

\newcommand{\bvec}[1]{\ensuremath{\mathbf{#1}}}
\def\bibsection{\section*{References}} 

\title{Wildcard error: Quantifying unmodeled errors in quantum processors}
\begin{abstract}
\noindent Error models for quantum computing processors describe their deviation from ideal behavior and predict the consequences in applications.  But those processors' experimental behavior -- the observed outcome statistics of quantum circuits -- are rarely consistent with error models, even in characterization experiments like randomized benchmarking (RB) or gate set tomography (GST), where the error model was specifically extracted from the data in question.  We show how to resolve these inconsistencies, and quantify the rate of unmodeled errors, by augmenting error models with a parameterized \textit{wildcard error model}.  Adding wildcard error to an error model relaxes and weakens its predictions in a controlled way. The amount of wildcard error required to restore consistency with data quantifies how much unmodeled error was observed, in a way that facilitates direct comparison to standard gate error rates. Using both simulated and experimental data, we show how to use wildcard error to reconcile error models derived from RB and GST experiments with inconsistent data, to capture non-Markovianity, and to quantify all of a processor's observed error.
\end{abstract}

\author{Robin Blume-Kohout}
\author{Kenneth Rudinger}
\author{Erik Nielsen}
\author{Timothy Proctor}
\author{Kevin Young}
\affiliation{Quantum Performance Laboratory, Sandia National Laboratories, Albuquerque, NM 87185 and Livermore, CA 94550}

\date{\today}
\maketitle

The performance of a quantum computing processor is constrained and defined by the errors it experiences.  Errors cause quantum circuits (programs) run on the processor to yield observed outcomes that deviate from their ideal, intended behavior.  Probabilistic \textit{error models} that describe these deviations are essential tools for assessing performance, predicting results of circuits, and informing error correction \cite{Johnson2017-xr, Trout2018-fy, Tuckett2018-bb, Baireuther2018-nz, Piedrafita2017-lp, Tuckett2019-ey, Bermudez2019-po, OBrien2017-fw, Gutierrez2013-ey, Magesan2013-ep, Florjanczyk2016-nt} and error mitigation \cite{Song2019-fg, Strikis2020-vm, Czarnik2020-zo, Hamilton2020-vj} strategies. Error models can be constructed from low-level physics descriptions~\cite{Debroy2019-pg,Bermudez2019-gw,Krantz2019-rk,McDermott2009-ng}, device characterization experiments such as randomized benchmarking (RB)~\cite{emerson2005scalable, knill2008randomized, magesan2011scalable, proctor2018direct} or gate set tomography (GST)~\cite{nielsen2020gate, blume2016certifying, dehollain2016optimization}, or simple heuristics. But even carefully constructed error models often yield predictions that are visibly inconsistent with experimental data. Inconsistencies are especially problematic when an error model derived from experiments disagrees with the training data used to construct it -- which, unfortunately, is common ~\cite{blume2016certifying, dehollain2016optimization, proctor2019detecting, wan2019quantum}. 

Inconsistency with experimental data makes it hard to take an error model seriously, or to trust its predictions for future experiments.  Including more potential error sources can help models fit their training data better, but this increased sophistication incurs increased experimental and analytical costs. And even complex models are unlikely to ever capture all observed errors. The alternative solution we pursue here is to rehabilitate simple error models, by quantifying \emph{unmodeled} errors and accounting for them in the model's predictions.  This is distinct from using standard statistical techniques based on hypothesis tests \cite{nielsen2020gate, rudinger2018probing} to detect statistically significant evidence of unmodeled errors, because those methods do not quantify the size or impact of unmodeled errors. If enough data is taken, even tiny unmodeled effects can generate signatures with overwhelming statistical significance!

\begin{figure}[t!]
  \centering
  \includegraphics[keepaspectratio=true,width=\columnwidth]{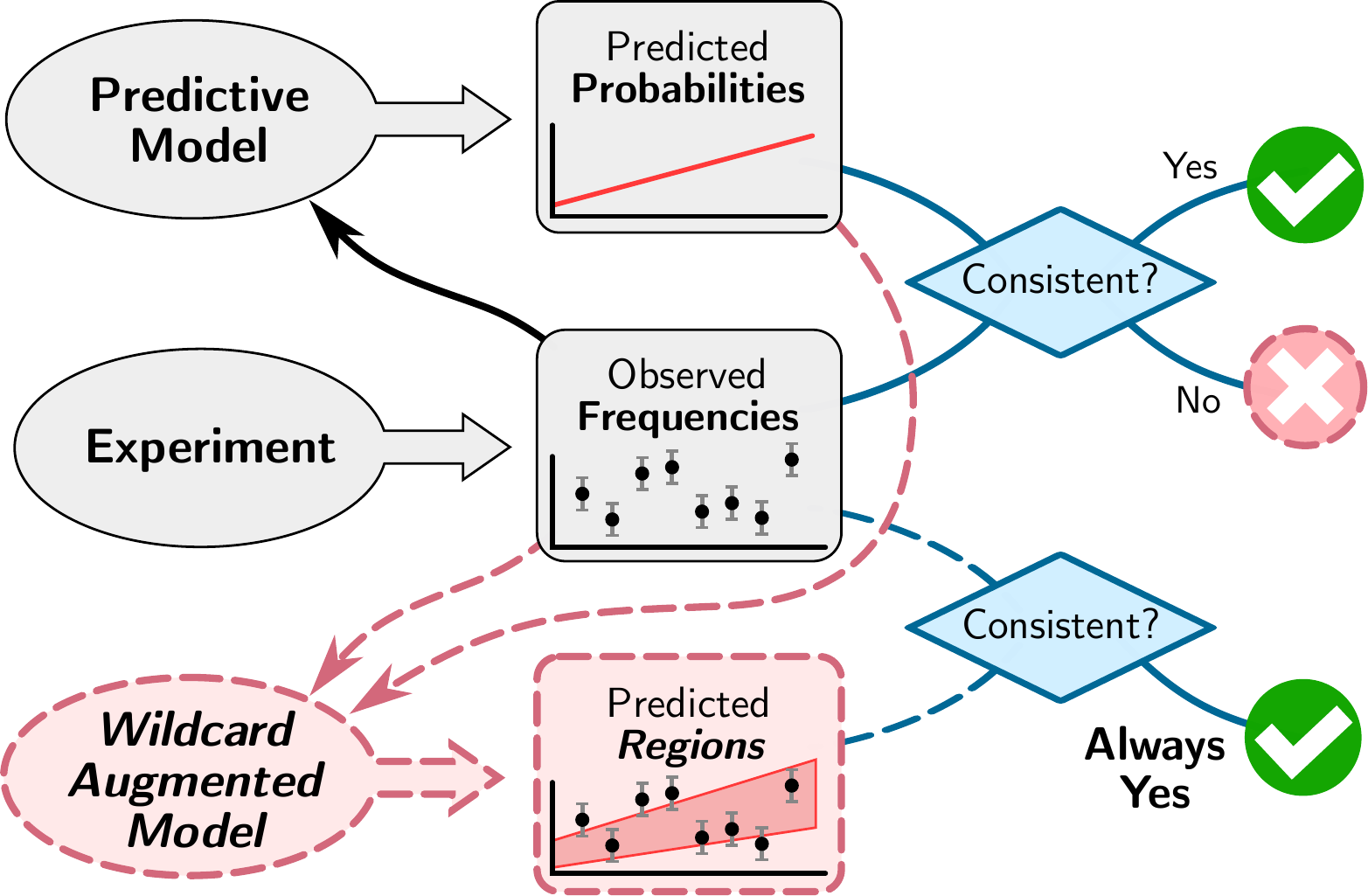}
  \caption{\textbf{Wildcard error models.} Error models for quantum processors are often inconsistent with experimental data, even when those data were used to build the model.  The inconsistency can be resolved by augmenting the original error model with a wildcard model that weakens its predictions, by expanding probabilistic predictions into regions.  The rates of errors missed by the original model can be quantified by the parameters of a minimal wildcard model that just barely makes the augmented model consistent with the data.}
  \label{fig:Flowchart}
\end{figure}

In this Letter, we introduce \emph{wildcard models}, a general technique to quantify unmodeled errors.  A wildcard model for a quantum processor augments an error model, weakening its predictions to make them consistent with experimental observations (see Fig.~\ref{fig:Flowchart}). As we demonstrate with simulated and experimental data, a ``wildcard error rate'' can be assigned to each of a processor's logic operations, to quantify the rate at which it induces unmodeled errors. Wildcard error rates can be used to quantify many different effects -- including non-Markovian  errors like leakage \cite{Wood2018-wi} -- by augmenting an appropriate error model.

\begin{figure}[t!]
\includegraphics[width=1\columnwidth]{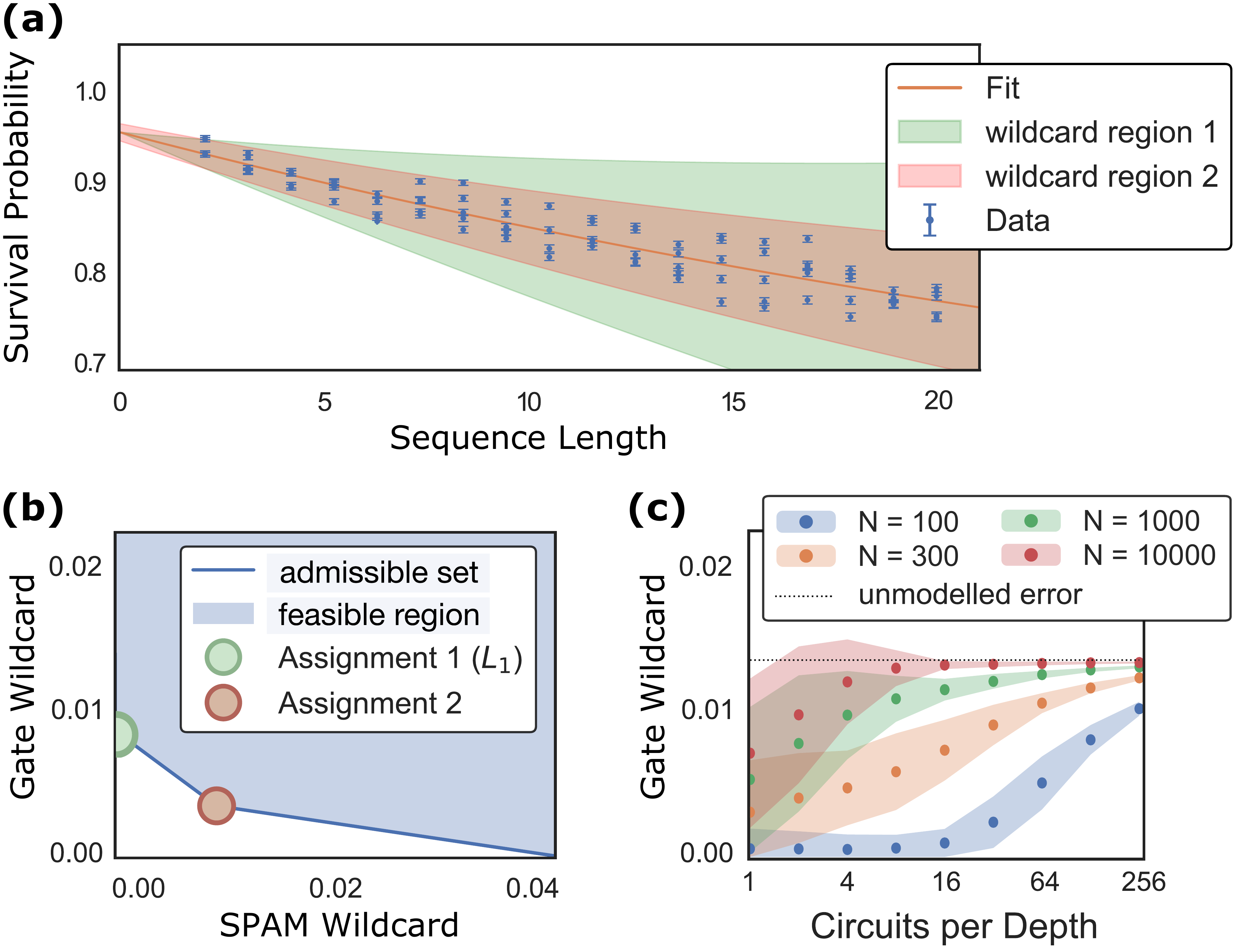}
\caption{\textbf{Quantifying unmodeled errors: a simple example.} 
Randomized benchmarking (RB) illustrates how wildcard can be used.  \textbf{(a)} Simulated data from individual RB circuits (points w/ $1\sigma$ error bars) on a qubit with $Z$-basis \emph{dephasing} errors are not consistent with a \emph{depolarizing} model (line) derived from RB data.  But augmenting that model with either of two minimal wildcard models (red/green bands) weakens its predictions and reconciles them with data. \textbf{(b)}  Wildcard models in (a) belong to a 2-parameter family assigning $\wildcard(C) = w_{\textsc{spam}} + d_Cw_{\mathrm{gate}}$, where $d_C$ is circuit $C$'s depth, and $\bvec{w}=(w_{\textsc{spam}},w_{\mathrm{gate}})$ are the parameters.  \emph{Feasible} models (blue region) reconcile every circuit $C$'s data with the error model's predictions, and are \emph{minimal} (blue line) if no strictly smaller model is feasible.  Red/green dots indicate the two minimal models in (a). \textbf{(c)} The average (dots) and 1$\sigma$ standard deviations (regions) of many simulations in which the number of RB circuits and the ``shots'' $N$ per circuit are varied, show that $w_{\mathrm{gate}}$ converges to $\frac12$ of the diamond distance ($\epsilon_{\diamond}$) between the true gates and the depolarizing model as $N\to\infty$.
}\label{fig:RB}
\end{figure}

\vspace{0.2cm}
\noindent\emph{Wildcard models.} For our purposes, an error model $\model$ for a quantum processor with logic operations $\{g_i\}$ is simply a map from circuits over $\{g_i\}$ (i.e., circuits composed of the gates in $\{g_i\}$) to probability distributions over their measurement outcomes. A model $\model$ predicts that running a circuit $C$ on the modeled quantum processor will result in a sample drawn from the outcome distribution $\model(C) = \vec{p}_C$. Many different error models are possible. Process matrices are a common and useful framework that can be learned from experiments~\cite{nielsen2020gate, blume2016certifying} or constructed from low-level physical modeling. But to introduce wildcard models we begin with a simpler alternative: the depolarizing model. In this model, each gate acts perfectly except that it induces uniform depolarization with probability $r$. The parameter $r$ can be estimated using RB. But if the gate errors are not depolarizing, the per-circuit RB data will typically not be consistent with a depolarizing model~\cite{Ball2016-ov}. To illustrate this, we simulated RB with single-qubit Clifford gates subject to $2\%$ $Z$-basis dephasing. Fig.~\ref{fig:RB}(a) shows the simulated data and the prediction of the depolarizing model estimated by standard RB analysis. There is clear statistical inconsistency between them. It can be quantified using a wildcard model.

We define a wildcard model $\wildcard$ for a quantum processor with logic operations $\{g_i\}$ to be a map that assigns a wildcard error $\wildcard(C) = w_C$ in the interval $[0,1]$ to each circuit $C$ over $\{g_i\}$. Combining a wildcard model $\wildcard$ and an error model $\model$ forms a wildcard-augmented model $\{\model, \wildcard\}$ that makes weaker predictions than $\model$. It predicts that the output distribution for the circuit $C$ is \emph{some} distribution in the ball $R_C$ of radius $w_C = \wildcard(C)$ around $\vec{p}_C=\model(C)$. The radius $w_C$ is measured in total variation distance (TVD), a metric that accurately captures the rate of incorrectly predicted events.  The TVD between two probability distributions $\vec{p}$ and $\vec{q}$ is $\delta_{\mathrm{TVD}}\left(\vec{p},\vec{q}\right) \equiv \frac{1}{2}\|\vec{p}-\vec{q}\, \|_1$. The region $R_C$ of distributions consistent with $\left\{\model,\wildcard\right\}$ therefore contains all and only those distributions $\vec{q}$ for which $\delta_{{\mathrm{TVD}}}\left(\vec{q},\vec{p}_C\right) \leq w_C$. A wildcard-augmented model is agnostic about which distribution within $R_C$ is correct. $R_C$ is not a distribution over distributions.

There are other contexts in which point predictions are relaxed to regions.  For example, uncertainty in an estimated model's parameters (``error bars''~\cite{Blume-Kohout2012-fq,Faist2016-pb}) can be represented as prediction regions~\cite{Geisser1993-ho}.  But that is distinct from what we do here with wildcard error -- those regions represent uncertainty about model parameters that could be reduced by taking more data, whereas wildcard models specifically represent \textit{out-of-model} effects.  These two types of prediction region could be combined, but we do not attempt to do so here.

\vspace{0.2cm}
\noindent \emph{Statistical consistency:} Wildcard models can weaken the predictions of an error model $\model$ so that they are consistent with observed data $D = \{\vec{f}_C\}_{C\in\mathbb{S}}$. Here, $\vec{f}_C$ is the empirical distribution from $N$ runs of the circuit $C$ and $\mathbb{S}$ is some set of circuits over $\{g_i\}$. As $N\to\infty$, probabilities $\vec{p}$ are consistent with data $\vec{f}$ if and only if $\vec{p}=\vec{f}$, so a wildcard-augmented model $\{\model, \wildcard\}$ is consistent with data $D = \{\vec{f}_C\}$ if and only if $\vec{f}_C \in R_C$ for all $C \in \mathbb{S}$, i.e.
\vspace{0.2cm}
\begin{equation}\label{eq:PerCircuitConstraints}
\delta_{\mathrm{TVD}}\left(\vec{p}_C, \vec{f}_C\right) \leq w_C \,\,\forall \, C \in \mathbb{S}.
\end{equation}
For finite $N$, statistical consistency has no unique definition. We choose to use a 95\% confidence hypothesis test that defines prediction regions $\{R_C\}$ and data $\{\vec{f}_C\}$ as \emph{inconsistent} if either \cite{rudinger2018probing}: (1) $\vec{f}_C$ rejects $R_C$ by a loglikelihood-ratio (LLR) hypothesis test with $\geq 97.5\%$ familywise confidence for any $C$
\footnote{When data used to fit the model are recycled for validation (e.g., assigning $\bvec{w}$), a small correction should in principle be made to the LLR test.  We ignore this correction, which is safe as long as the number of circuits is much great than the number of fit parameters.}, or (2) all the data together reject $\{R_C\}$ by an \emph{aggregated} LLR test at the $97.5\%$ confidence level.
Condition 1 captures effects with a large impact on a few circuits, while condition 2 captures effects with a small impact on many circuits. If a wildcard model satisfies these conditions we call it \emph{feasible}. In the RB example of Fig.~\ref{fig:RB}(a), many individual circuits' output statistics are inconsistent with the depolarizing model (by  condition 1), so a wildcard model is required to reconcile them.

\vspace{0.2cm}
\noindent \emph{Assigning wildcard error to each gate:} Structured wildcard models can quantify the ``amount'' of observed unmodeled error. To do so, we use wildcard models that assign a wildcard error rate $w_g \in [0,1]$ to each logic operation $g$ and then define $w_C$ as the sum over $C$'s component operations,
\begin{equation}
w_C = \sum_{g\in C}{w_g} = \sum_{g}{n_g(C) w_g} = \bvec{n}(C) \cdot\bvec{w},
\end{equation}
where $n_g(C)$ is the number of times that $g$ appears in $C$. This family of wildcard models is parameterized by a vector $\bvec{w} = \{w_g\}$ over $\{g_i\}$, so we denote a specific model in the family by $\wildcard_{\bvec{w}}$.  This ``wildcard error-per-gate'' parameterized family is simple, mirrors the general structure of many error models, and is faithful to the basic properties of imperfect quantum operations (e.g., subadditivity of worst-case error rates \cite{aharonov1998quantum}).

If $n_{g}(C) = n_{g'}(C)$ for all circuits $C\in\mathbb{S}$, then only the sum $w_g + w_{g'}$ impacts the per-circuit wildcard error $\{w_C\}$. In all examples herein, both state preparation and measurement appear once in every circuit, so only the sum of their wildcard error rates matters. To make this explicit, we write $\bvec{w}=(w_{\textsc{spam}}, w_{g_1}, w_{g_2}, \dots, w_{g_n})$ where $g_1$, $g_2$, $\dots$, $g_n$ are unitary logic gates. Sometimes it is useful to consider a restricted model that fixes certain gates' wildcard error rates to be equal. In our RB example (Fig.~\ref{fig:RB}), we assign each gate the same amount of wildcard error rate. This defines a family of wildcard models parameterized by $\bvec{w} = (w_{\textsc{spam}}, w_{\textrm{gate}})$ that assign $w_C = w_{\textsc{spam}}+ d_C w_{\textrm{gate}}$ to each circuit $C$ containing $d_C$ gates. Specific wildcard models in this family correspond to points in the $w_{\textsc{spam}} \times w_{\textrm{gate}}$ plane, as shown in Fig.~\ref{fig:RB}(b).

A useful feature of wildcard-per-gate models is that the condition for $\wildcard_{\bvec{w}}$ to be feasible --- i.e.,  that $\wildcard_{\bvec{w}}$ reconciles $\model$ with $D$ --- is computationally convenient. Since $w_C = \bvec{n}(C)\cdot\bvec{w}$, the $N\to\infty$ limit consistency condition of Eq.~\eqref{eq:PerCircuitConstraints} defines linear inequality constraints on $\bvec{w}$ that define a convex set of feasible wildcard models. Less obviously, the set of $\bvec{w}$ that satisfy the finite-$N$ statistical tests for consistency is also convex. The blue region in Fig.~\ref{fig:RB}(b) shows the convex set of feasible $\bvec{w}$ for the example shown in Fig.~\ref{fig:RB}(a).

\vspace{0.2cm}
\noindent \emph{Minimal wildcard models:} Augmenting an error model $\model$ with wildcard is an admission of that model's failure. So, we want to add a minimal wildcard model $\wildcard_{\bvec{w}}$ that weakens $\model$'s predictions only as much as necessary to restore consistency with the data. $\wildcard_{\bvec{w}}$ is minimal if there is no other feasible wildcard model $\wildcard_{\bvec{w}'}$ in our parameterized family for which $\wildcard_{\bvec{w}'}(C) < \wildcard_{\bvec{w}}(C)$ for some circuit and $\wildcard_{\bvec{w}'}(C) \leq \wildcard_{\bvec{w}}(C)$ for all circuits. Equivalently, $\wildcard_{\bvec{w}}$ is minimal if (and only if) there exists no feasible $\bvec{w'}\neq\bvec{w}$ for which $ w'_i \leq w_i$ for all $i$ and $w'_j<w_j$ for some $j$. Minimal wildcard models lie on the lower boundary of the convex set of feasible $\bvec{w}$, as shown (for our simple RB example) by the blue line in Fig.~\ref{fig:RB}(b).

Any minimal wildcard model is a plausible explanation of the data, so choosing among them requires secondary criteria. We usually select one that minimizes $\|\bvec{w}\|_1$. However, the minimizer of any convex  $f(\bvec{w})$ could be used instead.  We have investigated weighted linear functions; the red and green points in Fig.~\ref{fig:RB}(b) show $\bvec{w} = (w_{\textsc{spam}}, w_{\textrm{gate}})$ for two distinct minimal wildcard models that reconcile the simulated RB data with the depolarizing model.  Fig.~\ref{fig:RB}(a) shows the circuit success probability prediction regions of these two wildcard-augmented models (the green region is the $\|\bvec{w}\|_1$ minimum, which assigns $w_{\textsc{spam}}=0$). To understand what $w_{\textrm{gate}}$ quantifies in this example, we also repeated this simulation many times (using $\|\bvec{w}\|_1$ minimization), while varying the number of random Clifford circuits sampled at each depth ($K$) and samples of each circuit ($N$). As Fig.~\ref{fig:RB}(c) shows, the value of $w_{\mathrm{gate}}$ converges, as $N,K\to\infty$, to the half diamond norm distance ($\epsilon_{\diamond}$) between the gate process matrices of the true (dephasing) model and the estimated (depolarizing) model. We now explore and demonstrate some applications of wildcard error. 

\begin{figure}[t!]
\includegraphics[width=1\columnwidth]{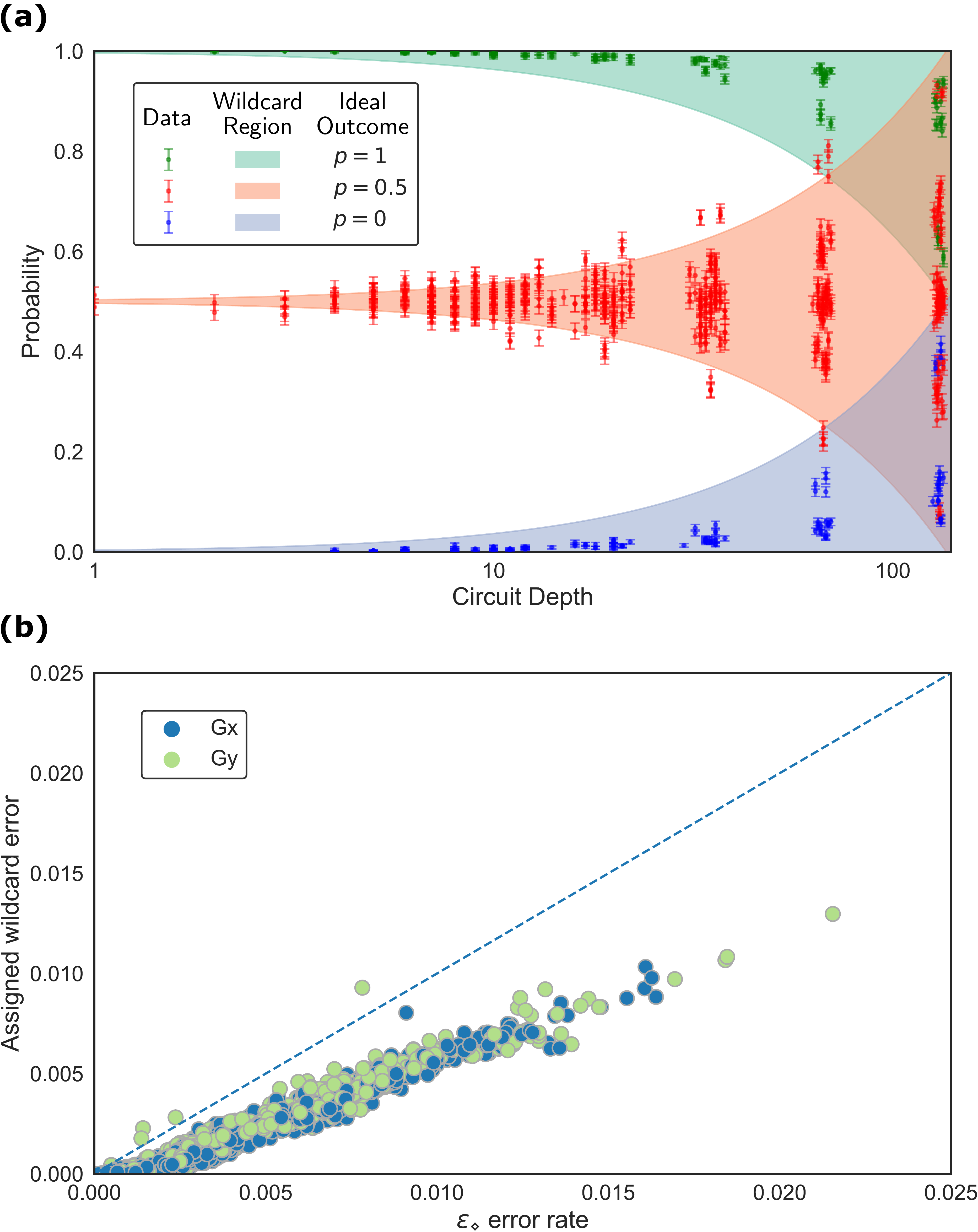}
\caption{\textbf{Quantifying \emph{all} error with wildcard}. The total observed rate of \emph{all} errors can be measured by augmenting the ``target'' model (error-free operations) with a wildcard model.  To demonstrate this, we simulated single-qubit GST circuits over two gates, $G_x$ and $G_y$. \textbf{(a)} One instance of simulated data (points, w/ $1\sigma$ error bars, colored according to ideal circuit outcome) versus circuit depth and prediction regions (similarly colored) of a wildcard-augmented model that relaxes the target model's predictions to make them consistent with the data. In this example, wildcard error rates for $G_x$ and $G_y$ were set equal ($w_x = w_y$) to facilitate visualization. \textbf{(b)} Results of 1000 independent simulations show that $w_x$ (blue) and $w_y$ (green) correlate well with the worst-case error rate ($\epsilon_{\diamond}$) of the $G_x$ and $G_y$ gates.  Each gate's error channel was chosen independently and randomly for each simulation, and $w_x$ was not set equal to $w_y$.}\label{fig:TotalError}
\end{figure}

\vspace{0.2cm}
\noindent \emph{Quantifying total observed error.} A simple, useful application for wildcard models is quantifying a processor's \textit{total} observed error rate. To do so, we augment a ``target'' error model consisting of the ideal gates, with a minimal wildcard model. Since each $\vec{p}_C$ is the ideal output distribution for $C$, the wildcard model is forced to explain all statistically significant deviations from ideal behaviour. Unlike a tomographic estimate of process matrices, such a minimal wildcard model will only quantify \emph{directly} observed error rates---so the results depend on what circuits were performed, as well as the underlying error rates. However, this analysis has advantages: it is much less computationally costly than most tomographic model fitting, it can be applied to data from any circuits, and it doesn't rely on the approximate correctness of a specific parameterized error model.

\begin{figure}[t!]
\includegraphics[width=1\columnwidth]{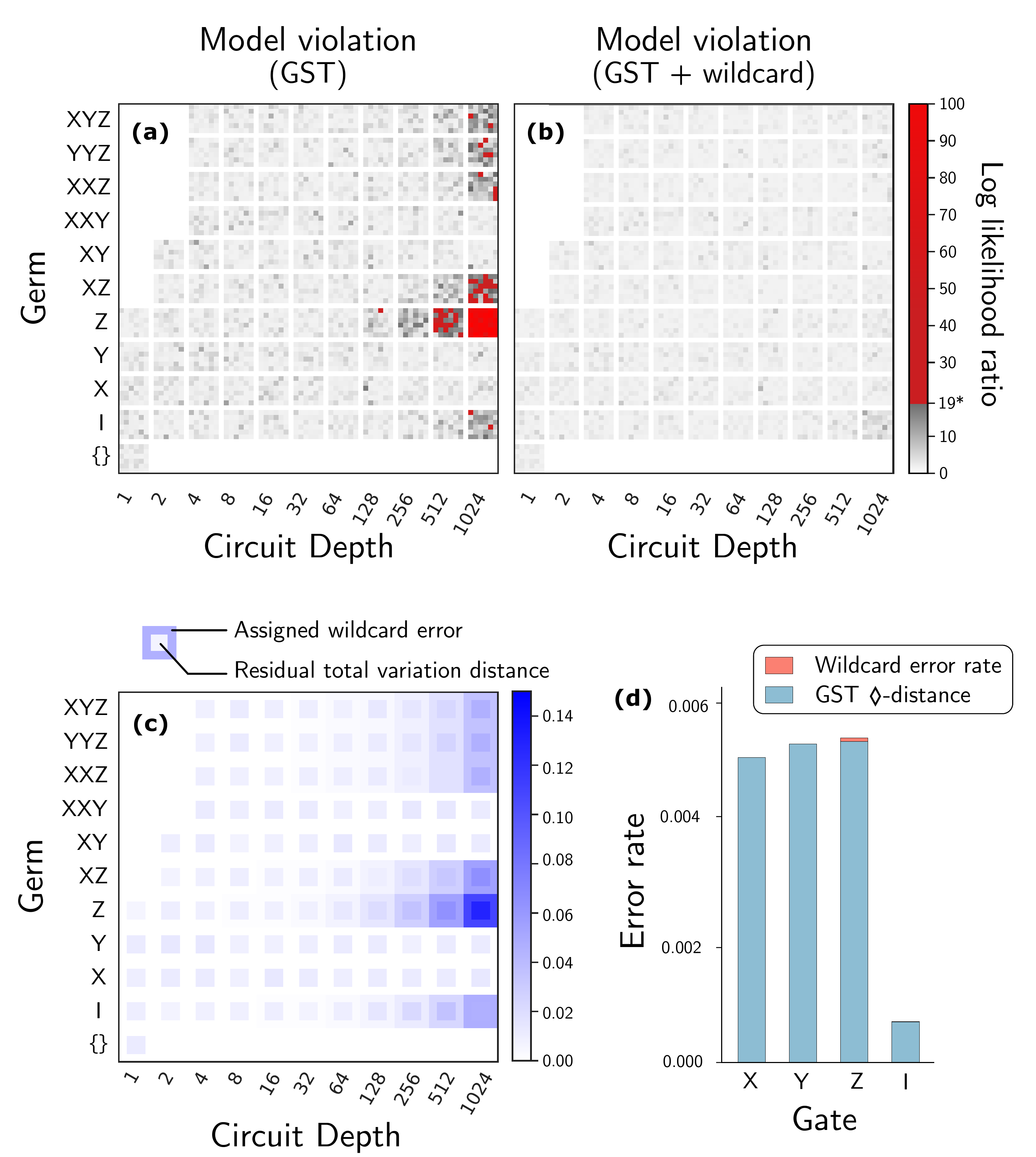}
\caption{\textbf{Wildcard captures non-Markovian errors in GST.} We simulated GST on a set of 4 gates ($G_x,G_y,G_z,G_i$) with non-Markovian errors induced by leakage on the $G_i$ ($10^{-4}$/gate) and $G_z$ ($3\times 10^{-4}$/gate) gates.  We used GST \cite{blume2016certifying,nielsen2020gate} to find a best Markovian (process matrix) fit, then augmented it with a 5-parameter wildcard model. Panels \textbf{(a,b,c)} show per-circuit data, arranged by the circuit's ``germ'' (vertical axis, labeled $\mathsf{X,Y,Z,I}$) and the circuit depth that results from repeating that germ (horizontal axis). \textbf{(a)} Loglikelihood ratios (LLRs) between the GST fit's predictions and data reveal large inconsistency (red boxes indicate statistically significant discrepancies). \textbf{(b)} LLRs between the wildcard-augmented GST model's predictions and the data show consistency.  \textbf{(c)} The amount of $w_C$ assigned to each circuit $C$ correlates well with the TVD between the GST fit's predictions and the observed frequencies. \textbf{(d)} Wildcard was only assigned to leaky operations, in amounts much less than the GST-estimated diamond-distance error of that gate.} \label{fig:Leakage}
\end{figure}

To demonstrate this technique (Fig. \ref{fig:TotalError}), we simulated data from single-qubit GST circuits \cite{nielsen2020probing}, that (collectively) have high sensitivity to all Markovian errors. We used a gate set containing $G_x$ and $G_y$ (where $G_p$ denotes a $\nicefrac{\pi}{2}$ rotation around the $\hat{p}$ axis), initialization to $\ket{0}$, and readout in the $Z$ basis.  Each operation was subject to depolarization, and the gates also suffer coherent over/under-rotation errors. Fig.~\ref{fig:TotalError}(a) shows  observed frequencies of ``1'' for each circuit versus circuit depth, from a simulation with $N=1000$ and where $G_x$ and $G_y$ are subject to the same over-rotation and depolarization rates. It also shows the prediction regions for a minimal wildcard model $\wildcard_\bvec{w}$ where $w_x$ and $w_y$ were fixed to be equal (to simplify visualizing these regions). Fig.~\ref{fig:TotalError}(b) summarizes the results of 1000 independent simulations with randomly chosen rotation and depolarization errors on the gates, comparing each wildcard error assignment to $\epsilon_{\diamond}$ for the corresponding gate used in that simulation. There is a strong correlation, although the wildcard error for each gate is consistently slightly less than $\epsilon_{\diamond}$ because although GST experiments are sensitive to worst-case error, tomographic analysis is required to infer it exactly. 

\vspace{0.2cm}
\noindent \emph{Quantifying non-Markovian error in GST:}  GST fits on experimental data often reveal statistically significant model violation \cite{blume2016certifying, dehollain2016optimization, proctor2019detecting}. Since $n$-qubit GST is designed to capture all $n$-qubit Markovian errors, this indicates non-Markovian errors that cannot be modeled by $n$-qubit process matrices. We could quantify these out-of-model errors by expanding the GST model to include other types of errors by, e.g., using a generic higher-dimensional model \cite{bennink2018quantum}, or modelling a specific effect like time-dependent gates \cite{proctor2019detecting} or leakage (which cannot be modeled as a single-qubit, completely-positive, trace-preserving process matrix). But larger models increase computational and experimental costs, and the errors outside of the GST model might be so small that they are of little practical importance. The easy alternative is to quantify the observed rates of non-Markovian errors in the GST data by augmenting the best-fit GST model with a minimal wildcard model. If the $\epsilon_{\diamond}$ error rates of the GST gate estimates are much larger than the wildcard errors assigned to the gates, then the GST process matrices represent a good model for the dominant errors.

To demonstrate this analysis we applied it to simulated GST data from a qubit that suffers leakage. We simulated the GST experiments on four gates: $G_i$, $G_x$, $G_y$, and $G_z$, where $G_i$ is an idle. All four gates were subject to Markovian errors \footnote{Consisting of depolarization and over-rotation errors with rates of $10^{-3}$ and $10^{-2}$, respectively.}, and the $G_i$ and $G_z$ gates also caused irreversible incoherent leakage at rates of $1 \times 10^{-4}$ and $3\times 10^{-4}$, respectively, to a state that is detected as $\ket{0}$. Without wildcard, the best-fit GST model is inconsistent with the data (Fig.~\ref{fig:Leakage}(a)). We constructed the  $\|\bvec{w}\|_1$-minimal wildcard model that reconciles this GST model with the data. It eliminates all of the inconsistency (Fig.~\ref{fig:Leakage}(b)) using only a tiny amount of wildcard error: $w_i = 1.1 \times 10^{-5}$, $w_z = 1.1\times 10^{-4}$, $w_x, w_y < 10^{-6}$ and $ w_{\textsc{spam}}=0$ (Fig.~\ref{fig:Leakage}(d)). The GST models' predictions are only weakened significantly for circuits that contain many $G_i$ or $G_z$ gates, and the amount of $w$ assigned to them correlates very closely with the TVD between the observed frequencies and the GST model's predictions (Fig.~\ref{fig:Leakage}(c)). The wildcard error rates assigned to $G_i$ and $G_z$ are lower than the underlying leakage rates, for two reasons. First, GST found process matrices that managed to explain \emph{some} of the leakage errors' effects.  Second, while GST circuits are sensitive to leakage, they are not designed for maximal sensitivity to it.  Neither these circuits nor the GST model are designed to distinguish leakage from Markovian errors. It is possible that \emph{jointly} fitting an error model and a wildcard model to data could mitigate the artifacts induced by out-of-model effects, but this remains an open question.

As a final test, we applied wildcard-augmented GST to two sets of experimental GST data from a 2017 trapped-ion experiment on the gates $G_i$, $G_x$ and $G_y$ \cite{blume2016certifying}.  The GST fits to these two sets of data exhibit model violation---over $30\sigma$ in the earlier experiment, and about $6\sigma$ in the last experiment.  This final experiment was described in Ref.~\cite{blume2016certifying} as remarkably Markovian, but even $6\sigma$ is overwhelming statistical evidence \cite{aad2012observation} -- so the errors in that experiment were certainly not perfectly Markovian! To quantify the observed non-Markovianity, we augmented each best-fit GST model with the $\|\bvec{w}\|_1$-minimal wildcard model.  The data from the early experiment could be reconciled by adding $w_{\textsc{spam}} = 0$, $w_i = 5\times 10^{-3}$, $w_x = 1\times 10^{-6}$ and $w_y = 1\times 10^{-4}$ to the GST estimate that had $\epsilon_{\diamond}$ error rates of $\epsilon_i = 7 \times 10^{-3}$, $\epsilon_x = 8 \times 10^{-5}$ and $\epsilon_y = 9 \times 10^{-6}$. In that experiment $w_y > \epsilon_y$ and $w_{i} \approx \epsilon_i$, implying that non-Markovian errors are a large proportion of the total error, and the GST estimate is likely to be unreliable. In contrast, the final experiment required only adding
$w_{\textsc{spam}} = 0$, $w_i = 2\times 10^{-6}$, $w_x = 7\times 10^{-8}$, and $w_y = 2\times 10^{-6}$ to a GST estimate with $\epsilon_{\diamond}$ error rates between $7 \times 10^{-5}$ and $1.3 \times 10^{-4}$.
 In this experiment, the smallest $\epsilon_{\diamond}$ was around $35\times$ larger than the largest wildcard error rate, so Markovian errors dominate and the GST estimate is reliable. This confirms and quantifies the ad hoc assertion in Ref.~\cite{blume2016certifying} that non-Markovian error in the final experiment is negligible compared with the Markovian errors, and it demonstrates that wildcard can be used to show that an error model captures most of the important behavior, despite being technically (and demonstrably) wrong.

\vspace{0.2cm}
\noindent\emph{Conclusions:}  Wildcard models can rehabilitate error models that are visibly wrong by qualifying their predictions, and wildcard error rates can be useful estimates of how much observed error wasn't captured by a given model.  Although we originally developed wildcard models to quantify non-Markovian errors in GST experiments, they can easily enhance other characterization protocols and other error models.  Specific types of error can be quantified by constructing an error model that \textit{doesn't} account for the chosen error, adding a wildcard model, and analyzing data designed to reveal the chosen error. Crosstalk is an excellent example, because although crosstalk errors are important for quantum computing \cite{proctor2018direct, rudinger2018probing, proctor2020measuring, harper2019efficient}, general crosstalk cannot be modelled efficiently \cite{sarovar2019detecting}.  The easy alternative is to build a crosstalk-free model \cite{sarovar2019detecting}, fit it to experiments that expose crosstalk, and use wildcard error rates to quantify the rate of crosstalk errors.  Wildcard analysis is intrinsically scalable, and because scalable error models must necessarily consider only restricted types of errors, we anticipate that wildcard will be especially valuable for many-qubit characterization protocols  \cite{harper2019efficient, erhard2019characterizing}. 

\begin{acknowledgments}
This work was supported by the U.S. Department of Energy, Office of Science, Office of Advanced Scientific Computing Research Quantum Testbed Program, and the Office of the Director of National Intelligence (ODNI), Intelligence Advanced Research Projects Activity (IARPA). Sandia National Laboratories is a multimission laboratory managed and operated by National Technology and Engineering Solutions of Sandia, LLC., a wholly owned subsidiary of Honeywell International, Inc., for the U.S. Department of Energy's National Nuclear Security Administration under contract DE-NA-0003525. All statements of fact, opinion or conclusions contained herein are those of the authors and should not be construed as representing the official views or policies of IARPA, the ODNI, the U.S. Department of Energy, or the U.S. Government.
\end{acknowledgments}

\bibliography{wildcard}

\end{document}